\documentclass[sn-mathphys-num,square, comma, numbers,sort&compress]{sn-jnl}

\usepackage{graphicx}%
\usepackage{multirow}%
\usepackage{amsmath,amssymb,amsfonts}%
\usepackage{amsthm}%
\usepackage{mathrsfs}%
\usepackage[title]{appendix}%
\usepackage{xcolor}%
\usepackage{textcomp}%
\usepackage{manyfoot}%
\usepackage{booktabs}%
\usepackage{algorithm}%
\usepackage{algorithmicx}%
\usepackage{algpseudocode}%
\usepackage{listings}%
\usepackage{amssymb,amsfonts,amsmath}
\usepackage{graphicx}      

\usepackage{amsthm}
\usepackage{listings}
\usepackage{textcomp}
\usepackage{mathtools}
\usepackage{dsfont}
\usepackage{color}
\usepackage{blindtext}%

\usepackage[utf8]{inputenc}
\usepackage[T1]{fontenc}
\usepackage{amssymb}
\usepackage{verbatim}
\usepackage{setspace}
\usepackage{xltabular}
\usepackage{booktabs}
\newtheorem*{pf}{Proof}
\newtheorem{thm}{Theorem}
\newtheorem*{nthm}{Theorem}
\newtheorem{rem}{Remark}
\newtheorem{lem}{Lemma}
\theoremstyle{definition}
\newtheorem{definition}{Definition}
\usepackage{setspace}
\usepackage{lineno}

\theoremstyle{thmstyleone}%
\newtheorem{theorem}{Theorem}
\newtheorem{proposition}[theorem]{Proposition}%

\theoremstyle{thmstyletwo}%
\newtheorem{example}{Example}%
\newtheorem{remark}{Remark}%
\newtheorem{corollary}{Corollary}%

\theoremstyle{thmstylethree}%

\begin{document}

\title[Article Title]{Oligopoly Game Stabilisation Through Multilayer Congestion Dynamics}


\author*[1]{\fnm{Toby} \sur{Willis}}\email{tewillis1@sheffield.ac.uk}

\author[1]{\fnm{Giuliano} \sur{Punzo}}\email{g.punzo@sheffield.ac.uk}
\equalcont{These authors contributed equally to this work.}


\affil*[1]{\orgdiv{Department of Automatic Control and Systems Engineering}, \orgname{University of Sheffield}, \orgaddress{\street{Street}, \city{Sheffield}, \postcode{100190}, \state{South Yorkshire}, \country{UK}}}




\abstract{		International trade and logistics are subject to factors including geopolitical instability, climate change, and black swan events such as the unforeseen closure of the Suez Canal. The problem of predicting local price change under modification of an underlying transport network or change in supply characteristics unites elements of game theory, network theory and transport. The Cournot Oligopoly models economic actors as rational players attempting to maximise profit by optimising supply quantities with analytical results now consolidated about equilibrium characteristics where transport conditions are fixed. Similarly, where supply and demand are fixed, the routing of goods in a transport network can be analytically solved through a traffic assignment problem. Hence we can solve the coupled Cournot-congestion problem by means of a 2-layer network. Where the layers are linked, inter-layer feedback wherein players attempt to maximise their utility occurs. In this respect we find players benefit from taking advantage of non-simultaneous responses to the market rather than moving to a new equilibrium. We draw conclusions about the nature of equilibria, finding that the concave utility curve property results in unique and stable equilibrium for each uncoupled layer, while linked layers have a non-unique stable equilibria for which general solutions are stated.}




\maketitle

\begin{xltabular}{\linewidth}{ l  X }
	\caption{Description of Variables} 
	\label{table: vardescription}\\
	\toprule
	Variable & Definition and Description  \\
	\midrule
	\endfirsthead
	\toprule
	Variable & Definition and Description  \\
	\midrule
	\endhead
	\bottomrule
	\endfoot
	
	
	$\mathcal{M}$ & The multilayer network \\
	
	$\mathcal{G}$ & The layers of the multilayer network \\
	
	$\mathcal{E_{\alpha\beta}}$ & The inter-layer edges of the multilayer network\\
	
	$G_\alpha$ & The upper layer, on which a Multimarket Cournot Game is played \\ 
	
	$G_\beta$ & The lower layer, on which there are congestion dynamics \\
	
	$x_i\in X$ & A goods seller and a player of the game across both layers \\
	
	$y_i\in Y$ & A market receiving goods from the multimarket Cournot game \\
	
	$N$ & The number of players \\
	
	$M$ & The number of markets \\
	
	$v_i^\alpha \in V_\alpha$ & A node in the upper layer. \\
	
	$e_{i,j}^\alpha \in E_\alpha$ & An edge between seller $x_i$ and market $y_j$ representing the amount of goods sold between one pair\\
	
	$v_i^\beta \in E_\beta$ & A node in the lower layer, representing a physical location\\
	
	$e_{i,j}^\beta \in E_\beta$ & An edge between $v_i^\beta$ and $v_j^\beta$\\
	
	$p_{i,j}^k$ & The k$^th$ set of edges in the lower layer between $v_i^\beta$ and $v_j^\beta$ \\
	
	$R$ & The reserve price function, giving sale price per good as a function of goods sold to the market \\
	
	$s_i \in S_i$ & The strategy of player $x_i$\\
	
	$S_{-i}$ & The strategy of all players except $x_i$\\
	
	$A_i(S) $ & The profit made by player $x_i$ in the upper layer\\
	
	$B_i(S)$ & The loss made by player $x_i$ in the lower layer\\
	
	$a_{i,j}(\cdot)$ & The profit function for player $i$ in market $j$\\
	
	$b_{i,j}(\cdot)$ & The congestion function for edge $e_{ij}$ in the lower layer\\
	
	$Q$ & The profit per good in a market when there are no sales\\
	
	$u(x_i,t)$ & The utility function for player $x_i$ at time $t$\\
	
	$t_h$ & The time horizon that players are optimising for. They consider utility before and including this time, but not afterwards \\
	
	$g$ & The profit at time 0 for player $x_i$ \\
	
	$s_n$ & The sequence of profits made by player $x_1$ \\
	
	$s_P$ & A pareto optimal strategy profile \\
	
	$T$ & The fixed total transport \\
	
	$f(\cdot)$ & A function giving the amount of transport on an edge \\
	
	$P_{i,j}$ & A mapping from $e_{i,j}^\alpha$ in the upper layer to each set of edges in the lower layer which are a transport solution for that edge \\
	
	$\mu_{i,j}$ & The set of edges in the upper layer with use edge $e_{i,j}^\beta$\\
	
	$\Lambda$ & A set of $\lambda_{\dots}$ which gives the amount of edges which transport each type of traffic\\
	
	$G_{\beta}^k$ & The k$^{\mbox{th}}$ subgraph of $G_\beta$\\
	
	$f_i(e_j)$ & The amount of flow player $i$ uses on edge $e_j$.\\
	
	$c_i(\cdot)$ & The cost of market saturation\\
	
	$l(e_j)$ & The length of edge $e_j$ \\

\end{xltabular}

\section{Introduction}\label{sec:introduction}

Initially designed for modeling duopolistic scenarios, the Cournot competition model is an economic framework that is applicable to oligopolistic situations where multiple firms compete in the same markets for products, such as oil \cite{OPEC} or grain \cite{Grain}. In Cournot competitions, a player's rewards from selling goods are contingent on the actions of other players, as these actions influence the selling price of such goods. Players strive to maximize their profits through their strategic choices, and their counterparts select strategies in response, as illustrated in the Cournot model \cite{Cournot1897}. Cournot competitions are often referred to as Cournot games due to their strategic nature. These games model markets by establishing a relationship between the quantity of goods produced and the resulting prices. In terms of market dynamics, a considerable body of research has focused on single-market Cournot games. The literature, \cite{Varian_advanced}, \cite{Mas_Collel} has yielded insights into the characteristics of equilibria in these contexts. 
\cite{Bimpikis2} introduced a unique approach by partitioning the players and the market, which resulted in equilibria for Cournot competition. In this novel framework, players are linked to particular markets through a bipartite network. \cite{KYPARISIS1990}, \cite{Qiu}, and \cite{Abolhassani2014} independently achieved results in this area, wherein they defined an equilibrium allocation as a collection of strategies in which companies realize zero marginal profit. Notably, Cournot competitions do not incorporate factors related to variable transportation costs or consider the physical distances between players and the markets in which they engage in competition.

The widely recognized selfish routing problem takes into account the expenses associated with traversing physical distances. In this context, players navigate a network from a start to a destination node. In this network, each edge incurs a cost, which is influenced by the volume of players utilizing it. As more individuals opt for the same edge, its cost increases, thereby reducing the overall benefit or utility experienced by the players \cite{Roughgarden2006}. Within selfish routing problems, a Wardrop equilibrium is characterized by the collection of routing strategies employed by players, where no player has any motivation to independently alter their chosen route. This equilibrium concept plays a pivotal role in the traffic assignment problem, to the extent that observed flows are frequently regarded as the naturally arising equilibrium and serve as an initial basis for deriving the origin-destination matrix. This is accomplished by reconstructing the costs linked to each edge through congestion functions \cite{Zhang2019}. Congestion functions are mathematical representations that very often exhibit convexity and monotonic increase. They establish a relationship between the travel time along a specific road segment (often used as a proxy for cost) and the volume of traffic present on that segment. While distinct from Cournot games, selfish routing problems have also been explored within the context of multilayer networks. In these scenarios, each layer of the network corresponds to different modes of transportation or performance attributes, such as long-range, high-speed journeys in contrast to shorter-range, slower transfers. In the realm of multilayer transportation systems, \cite{Ibrahim2021} conducted an investigation and offered algorithms designed to compute Wardrop equilibria, thereby addressing the equilibrium solutions for this complex transport setting.

A multilayer setting lends itself well to the representation of different dynamic systems, with each system abstracted as a network and interacting with others. Specifically, exploration of network Cournot competition \cite{Bimpikis} can be integrated with a separate transport layer, where the selfish routing problem concerning the transportation of goods from players to markets is investigated.

In this context, it becomes necessary to consider the Nash equilibrium of the Cournot game in conjunction with the Wardrop equilibrium of the selfish routing problem, as either or both of these equilibria may be achieved. Under certain appropriate assumptions, it's worth noting that where a Nash-Cournot equilibrium is defined and conditions for existence and uniqueness of it are given, then the asymptotic behavior of this equilibrium is shown to yield a total flow vector corresponding to a Wardrop equilibrium \cite{haurie1985relationship}. This corresponds to the similarity between traffic assignment and Cournot where player market-share is infinitesimal.

The first layer of the two layer model introduces a Cournot competition conducted within a bipartite network involving players and markets, while the second layer presents a selfish routing problem akin to a congestion game played on the physically embedded transportation network.

Results about the nature of linked oligopoly-transport system are given in \cite{Alsabah2021} which delves into a capacity-constrained oligopoly problem and identifies that a reduction in transportation costs can have a detrimental effect on the profits of all firms involved. It's important to note that their approach differs from the one presented in this paper, as transportation within a network is not considered. Transportation is instead assigned a functional cost.

In this study, we build upon the Cournot competition framework introduced in \cite{Bimpikis} and leverage the insights from \cite{Ibrahim2021} for the congestion game aspect. The synergy between these two components yields original outcomes, particularly in the emergence of equilibrium points where individual players lack any incentive to alter the quantities of goods they sell in each market or the routing choices for shipping these goods. It is worth mentioning that differently from \cite{WILLIS2022}, results about optimal response under different patience characteristics, the effect of player collusion and an analysis of numerical results on equilibrium formation in a 3-player game are given.

While both network Cournot cases and transportation problems are well-documented topics in the literature, there exists a notable void concerning the interplay between these issues. This paper fills that gap by exploring the dynamics of the bilayer model.

In this paper, we present the following original contributions: 
\begin{itemize}
	\item We build on existing network oligopoly models e.g. \cite{Bimpikis}, and develop them through the introduction of a second layer following congestion based transport dynamics. While these dynamics rely on classical network congestion games, for example, \cite{Wardrop}, their effect is to change the oligopoly competition in unexplored ways. The market prices are influenced by the cost of transport to those markets. The volume of traffic through the transport layer is influenced by the market conditions.
	\item We establish the existence and uniqueness of equilibrium points within each layer, considering the hypothesis of stationary conditions in the other layer. Additionally, we demonstrate the stability of these equilibrium points individually within each layer for fixed total production.
	\item We conduct an analysis of the coupled co-evolving dynamics between both layers. Specifically, we identify the equilibrium points that emerge from this interplay and provide rigorous proofs of their uniqueness and stability. This analysis sheds light on the intricate dynamics of the interconnected layers in our model.
\end{itemize}

This study primarily centers on the analytical facets of the Cournot-congestion game. However, it also delves into applied research inquiries related to the intricate connection between market competition and transportation expenses. This theme holds significant practical significance, influencing strategic decisions related to crucial areas like food security and supply chain resilience \cite{Nagurney}.

The transition from the theoretical model to real-world applications is facilitated by conceiving the global market as a routing problem, where firms act as competing agents striving to optimize their individual utility. This conceptual bridge underscores the relevance and broader implications of our research.

\color{black}
%
%
%

\section{Model Formulation}
\label{Section:Model_Formulation}
\subsection{Mode and Network Formulation}
The bi-layer model captures the impact on the price in the markets in the upper layer of transport through the lower layer. Players sell good attempting to maximise their profit, evaluating with respect to both the price of goods in the various markets and the cost of transporting goods to those markets. The players attempt to maximise their total profit, captured by the costs of transporting goods and the profit from selling to the various markets.

\begin{figure}[h]
	\centerline{\includegraphics[width=72mm]{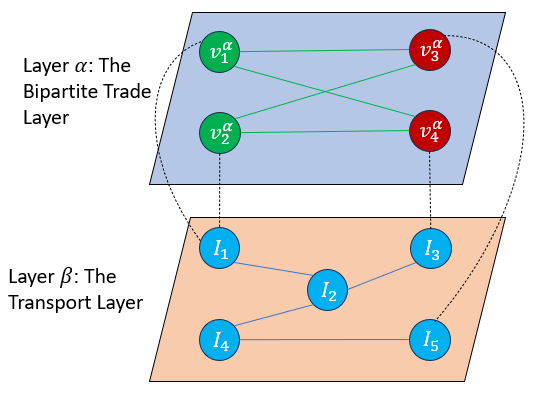}}
	\caption{An example model with 2 players, 2 markets and a 5 node transport layer}
	\label{3dmultilayer2}
\end{figure}

\color{black}

We shall consider a multilayer network with two distinct layers, referred to as a bilayer network. The formulation follows the multilayer convention established in \cite{Boccaletti2014}.


A bilayer network is a pair represented as:
\begin{equation}
	\mathcal{M}=(\mathcal{G},\mathcal{E}_{\alpha\beta})
\end{equation}
Here, the indices $\alpha$ and $\beta$ are used to reference the layers, and the family of undirected weighted graphs denoted as $\mathcal{G}=(G_\alpha,G_\beta)$ is referred to as the ``layers" of the network $\mathcal{M}$. In the context of Figure \ref{3dmultilayer2}, the graph $G_\alpha$ is also colloquially called the ``Upper Layer," while $G_\beta$ is referred to as the ``Lower Layer".

Layer $\alpha$ is the graph given by 
$G_\alpha=\left({V_\alpha},{E_\alpha}\right)$ 
where $V_{\alpha}=\{v_i^{\alpha}, i=1,2\dots N_{\alpha}\}$ with $N_\alpha$ the number of nodes in layer $\alpha$. $N_\alpha=N+M$ where $N$ is the number of players (sellers) and where $M$ is the number of markets.

$
a_{i,j}=(v_i^{\alpha},v_j^\alpha)\in E_\alpha
$
where $E_\alpha$ is the set of intra-edges and $a_{i,j}$ is a single edge. With $v_i^\alpha\in X$ and $v_j^\alpha \in Y$. 

The upper layer's nodes can be partitioned into two sets: $X$, consisting of $N$ players (sellers), and $Y$, consisting of $M$ markets.

\noindent A bipartite assumption is appropriate since there is no exchange of traffic between groups of sellers or among groups of markets.

\noindent As layer $\alpha$ is bipartite, the following relations are true.
$$
X\cap Y=\emptyset
$$
and
$$
V_\alpha = X \cup Y.
$$


An ordering exists on the elements of $V_\alpha$ with $v_1^\alpha=x_1,\dots,v^\alpha_N=x_N,v^\alpha_{N+1}=y_1\,\dots,v^\alpha_{N_\alpha}=y_M$.
Similarly, in layer $\beta$ we have the (non-bipartite) graph 
$
G_\beta=\left({V_\beta},{E_\beta}\right),
$
where
$
V_{\beta}=\{v_i^{\beta}, i=1,2\dots N_{\beta}\}
$
and
$
b_{i,j}=(v_i^\beta,v_j^\beta)\in E_\beta
$
with $N_\beta$ the number of nodes in layer $\beta$.

Each node on the upper layer (and accordingly every member of these sets) is associated to at most one node in $G_\beta$. The discrete `location' map  \cite{WILLIS2022}
\begin{equation}
	L(\cdot)=v_i^{\beta}:\mathbf{V_\alpha}\rightarrow\mathbf{V_\beta}
\end{equation}
gives the geographical embedding of any player or market onto $G_\beta$ and for each $v_k^\alpha \in G_\alpha$, there exists  $e_k\in \mathcal{E}_{\alpha\beta} = \{e_k:e_k=(v_k^{\alpha}, L(v_k^{\alpha}))\}$. $L(\cdot)$ is not an injective function so there is no limitation requiring players and markets to be in different nodes.

It has been given that $E_{\alpha}$ and $E_{\beta}$ are the sets of intra-layer edges.
$
\mathcal{E}_{\alpha\beta}
$
is defined to be the set of inter-layer edges between layers $\alpha$ and $\beta$.

The weights of the edges in $E_\alpha$ given by 
\begin{equation}
	w(\cdot): E_\alpha \rightarrow \mathbb{R}_+
\end{equation} 
represent the amount of goods being sold by a player to a market.



Transportation dynamics appear within the lower layer $G_\beta$, with its nodes and edges representing a real world transportation system. The edges in the lower layer have lengths $l:E_\beta \rightarrow \mathbb{R}_+$and capacities, $c:E_\beta \rightarrow \mathbb{R}_+$.


Finally, the set $\mathcal{E_{\alpha\beta}}$ of inter-layer edges represents the geographical embedding of the elements of the upper layer into the lower layer. 


\subsection{Transport Routing}
\label{Sec:Paths}

The implementation of transport paths which appears below follows the implementation appearing in \cite{WILLIS2022}.

Transport on the lower layer must satisfy the demand in the upper layer. In other words, the quantity of goods transferred between a player and a market in the upper layer must correspond to the travel demand between the respective player and market locations in the transport layer. While there is only a single path between $x_i$ and $y_j$ in the upper layer, there can be multiple paths between $L(x_i)$ and $L(y_j)$ in the lower layer. Each feasible route is denoted as a path $p_{i,j}^k$, which represents an ordered sequence of edges in the lower layer. Here, $i$ and $j$ signify the indices of the elements in sets $X$ and $Y$ that the path connects, and $k$ serves as the path index.

The set of all paths between $L(x_i^\alpha)$ and $L(y_j^\alpha)$ is referred to as $P_{i,j}$, and the collection of all path sets between $L(X)$ and $L(Y)$ in the lower layer is defined as $P$. This is structured as follows:
\begin{equation}
	P=\left(P_{1,1},\dots,P_{1,M},P_{2,1},\dots,P_{N,1},\dots,P_{N,M}\right)
\end{equation}
and
$$
P_{i,j}=\left(p_{i,j}^1,p_{i,j}^2,\dots,p_{i,j}^{N_{P_{i,j}}}\right),
$$
where $N_{P_{i,j}}$ signifies the count of paths in the lower layer between node $L(x_i^\alpha)$ and $L(y_j^\alpha)$.

The transportation through the lower layer must adhere to the transportation requirements as described by the edge flow in the upper layer. Consequently,
\begin{equation}
	f(a_{i,j})=\sum_{k=1}^{N_{P_{i,j}}} p_{i,j}^k.
\end{equation}

Note that, with a slight abuse of notation, we've defined $p_{i,j}^k$ to represent both the path (an ordered set of edges between $L(x_i^\alpha)$ and $L(y_j^\alpha)$) and the flow of goods routed through them.

\subsection{The Cournot Oligopoly Game}
\label{Sec:Cournot}
A Cournot competition, as originally outlined in \cite{Cournot1897} describes the dynamics of the players in the upper layer with respect to each market as the following conditions are satisfied:
\begin{itemize}
	\item The game has multiple players;
	\item there is no collusion between players;
	\item each player has market power, that is, each player strategy changes the price in the market they participate in;
	\item The number of players and markets are fixed and the goods sold by all players are homogeneous;
	\item players choose the quantity of goods to sell rather than their price, which is a consequence of the total amount of good sold;
	\item players engage in rational behaviour.
\end{itemize}
The condition of fixed total production, with players assigning goods to each market results in players engaging in parallel Cournot competition across the markets.

\begin{figure}[h]
	\centerline{\includegraphics[width=0.5\columnwidth]{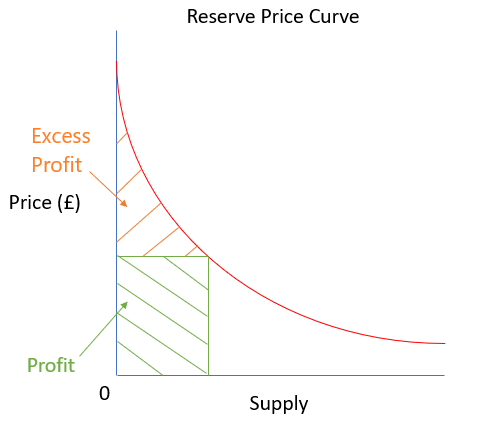}}
	\caption{A sample reserve price curve with profit in green and excess profit in orange}
	\label{Reserve_Graph}
\end{figure}

A reserve price (or the indifference price) is the minimum price a market is willing to pay for a good. In each market it is assumed there will be a range of reserve prices held by the spectrum of buyers. This leads to a functional relationship between the supply of goods and the price at which they will all be sold.
This will be the same price for all buyers, meaning that buyers potentially willing to pay more for the same goods will be better off. 
Associated with each market $y_j$ is a reserve price function $A_{i,y_j}$ of the form
$$
A_{i,y_j}(\cdot): \mathbb{R}_+ \rightarrow \mathbb{R}
$$
from a positive supply to a profit (which could be negative). As supply increases price will be monotonically decreasing. Players seek to maximise their profit by splitting their sales between the available markets in order to receive maximal utility. The relationship between supply and price per unit can be seen in figure~\ref{Reserve_Graph}.

\subsection{Game Dynamics}

Every player, denoted as $i$, strives to optimize the difference between their income and the costs incurred in selling their goods. This multilayer game, originally presented \cite{WILLIS2022} appears here for clarity. The key aspects of this game include:

The costs incurred relate to the utilization of transport links, and these costs increase as the transport links become more congested due to higher usage.

Income is derived from selling goods to markets, and players earn profit based on the supply and demand dynamics in the markets they sell to.

The utility function for each player is defined as follows:
\begin{equation}
	u(s_i, s_{-i}): S_i \times S_{-i} \rightarrow \mathbb{R}_+.
\end{equation}

Here, $s_i$ represents the current strategy of player $x_i$, $S_i$ is the set of all possible strategies for player $x_i$, $s_{-i}$ is the current strategy of all other players, and $S_{-i}$ is the set of all possible strategies for all players.

This is a dynamic game with players moving asynchronously. Players sequentially update their strategies to optimize their responses to the strategies chosen by all other players. Players consider not just the immediate future but seek to maximize their utility over multiple rounds, taking into account the future strategies of other players. The update order starts with player 1 (located at $x_1$), followed by player 2, and so on, looping back to player 1 after player $N$ has updated their strategy.

Each player's strategy is a combination of actions from both an upper layer and a lower layer, which includes:
\begin{enumerate}
	\item Choosing an allocation of markets to sell goods in the upper layer. This involves a mixed strategy in the form of $(w(a_{i,1}), w(a_{i,2}), \dots, w(a_{i,M}))$ for $a_{i,j} \in E_\alpha$.
	
	\item Selecting a mixed strategy for the paths they will use to transport goods, ensuring they fulfill their origin-destination pairings.
\end{enumerate}
This results in a strategy represented as:
\begin{equation}
	s_i=\begin{pmatrix}
		[a_{i,1},a_{i,2},\dots,a_{i,M}]\
		[p_{i,1}^1,\dots,p_{i,1}^{(N_{P_{i,1}})},\dots,p_{i,m}^1,\dots,p_{i,N_c}^{(N_{P_{i,m}})}]
	\end{pmatrix}
	\label{EquationStrategy}
\end{equation}

The utility for a player $x_i$ is calculated as:
\begin{equation}
	u_i(s_i, s_{-i})=A_i(s_i, s_{-i})-B_i(s_i, s_{-i})
	\label{Equ:Utility}
\end{equation}
\begin{equation}
	A_i(s_i, s_{-i})=\sum_{j=1}^{M} A_{i,m_j}(s_i, s_{-i})
	\label{Equ:A2}
\end{equation}

Here, $A_i(s_i, s_{-i})$ represents the total profit made in the upper layer by player $i$, and $A_{i,m_j}(s_i, s_{-i})$ denotes the profit made by player $i$ in market $j$. $B_i(s_i, s_{-i})$ accounts for the congestion cost paid by player $i$. The choice of congestion functions is not restricted beyond requiring a monotone increase, continuity and convexity.

\section{Time and Player Independence}

Consider a group of players aiming to profit from various markets. They receive updates on the current market state sequentially and aim to maximize their utility. The cost of market saturation $c_i$ for each player $i$ is determined as follows in a 2-player 2-market game:
\begin{equation}
	c_i=f_i(e_1) \left(f_i(e_1) + \sum_{j\neq i}^{N}f_j(e_1)\right) +(1-f_i(e_1)) \left((1-f_i(e_1))+(N-1)-\sum_{j\neq i}^{N}f_j(e_1)\right)
	\label{A3}
\end{equation}

In this equation, $f_i(e_k)$ represents the amount of traffic that player $i$ assigns to route $k$, and $N$ is the total number of players. The choice of an inverse linear demand function, as expressed in equation~\ref{A3}, is a commonly used assumption in the literature \cite{Vives} \cite{Singh}. Extension to additional players and markets requires an expansion of this equation. This results in the general form
$$
c_i=\sum_i=1^M f_i(e_1)\left(\sum_j^N f_j(e_1) \right).
$$

\color{black}

\begin{rem}
	This model is time independent and is symmetric to all players. As such, all players will follow the same strategy as on their turn they consider the same problem.
\end{rem}

Players update their responses sequentially beginning with player 1, continuing to player $N$ and returning to player 1 afterwards. Beginning with player $1$, the situation before player $1$ changes strategy is considered to be the set of strategies at time $t=0$. Accordingly at the time step $t_k$, the player $i$ has just updated where $t_i=i (mod N)$. 

As player $i$ gets to update their strategy again at time step $t_k+N$ they are only interested in their received utility at time steps $t_k, t_k+1,\dots,t_k+(N-1)$. Without loss of generality we can consider the player who is about to respond to be player $1$ at $t=0$. Given a starting situation $[-,s_2,s_3,\dots,s_N]$ player $1$ wants to choose the strategy $s_i$ which minimises their costs at time steps $1,2,\dots,N$.

\section{Uncoupled Dynamics Results}

We shall first consider the characteristics of the layers when examined individually (such that there is no feedback from the other layer) to then understand the behaviour of the model as a whole.
\subsection{Upper Layer Dynamics - Approach to Cournot Equilibrium}
\label{Upperlayeronly}
To analyse the dynamics within the upper layer, let's begin by establishing an assumption that will later be relaxed: the allocation of flows across all paths in the lower layer connecting the same origin and destination remains fixed. In other words, for all $k$ and $l$ within the range of $1$ to $N_{p_{ij}}$, and for all $i$ and $j$ within the range of $1$ to $N$ and $1$ to $M$ respectively, it holds that $p_{ij}^k=p_{ij}^l$. Furthermore, this allocation remains constant and unaffected by the quantity of goods a player sells in a specific market.

This situation means that while the transportation costs have an impact on a player's utility, the dynamics in the lower layer are unresponsive to changes in the upper layer. In essence, a player can adjust the volume of goods they sell in each market, but they lack the ability to alter the distribution of flow among paths with the same origin and destination. As a result, the game unfolds exclusively in the upper layer, although it is influenced by the costs generated in the lower layer.

If we assume that the game commences from a non-equilibrium state and the players progressively respond optimally to the strategies of other players, we can analyze the utilities over multiple rounds.

\begin{figure}[h]
	\centering
	\begin{minipage}{0.45\textwidth}
		\centering
		\centerline{\includegraphics[width=\columnwidth]{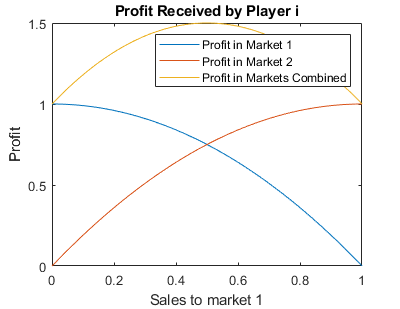}}
		\caption{The upper layer utility functions for player $i$}
		\label{Utility_Graph}
	\end{minipage}\hfill
	\begin{minipage}{0.45\textwidth}
		\centerline{\includegraphics[width=\columnwidth]{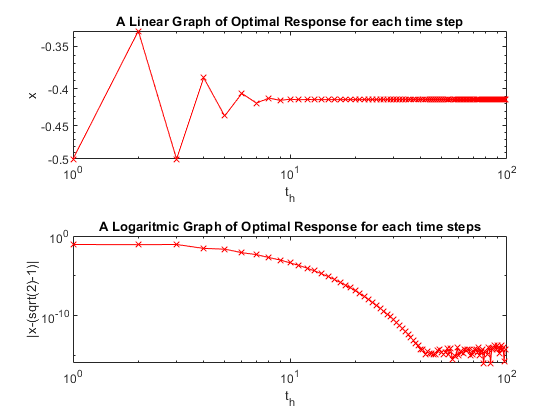}}
		\caption{The set of $x$ corresponding to the optimal response $(x) (a_0)$ for $1\leq t_h\leq 1000$, analytic convergence is proved in \ref{Thm:time_horizon_4} so the unusual behaviour beginning at $t_h=40$ is due to numerical limitations.}
		\label{Fig:timehorizonmedium}
	\end{minipage}
\end{figure}

The Cournot Game has a unique equilibrium where the utility function is twice differentiable, concave, and strictly decreasing and the costs of production are twice differentiable, convex and increasing \cite[][Theorem 1]{Bimpikis}. 
The assumption is made that there is no cost for production in the upper layer.
As such, this cost function is given by the 0-function which satisfies the conditions outlined in theorem 1 in \cite{Bimpikis}. 
Consider the two player two markets case with a profit function $A_{i,m_j}(s_i,s_{-i})$ generally defined as 

\begin{equation}
	A_{i,m_j}(s_i,s_{-i})=w(a_{i,j}) \left(Q-\sum_{k=1}^{N} w(a_{k,j}) \right),
	\label{Equ:a}
\end{equation}
where $Q$ is a fixed constant giving the profit where there is no supply. As the players aim to maximize their utility without the ability to adjust their production, it's important to note that the magnitude of $Q$ does not impact the analytical outcomes. The utility that an individual player receives in a market adheres to the curve labeled 'profit in market 1,' which is depicted in Figure~\ref{Utility_Graph}.
A profit function, which satisfies the conditions outlined by \cite{Bimpikis}, does exist. This implies that the game possesses a unique equilibrium, as demonstrated in their work.

To identify this equilibrium, we can examine the sum of $A_{i,m_1}(f(a_{1,1}),s_{-i})$ and $A_{i,m_2}(1-f(a_{1,1}),s_{-i})$ for a variable $h$ within the range of [0,1]. Here, $m_1$ and $m_2$ represent markets, and $f(a_{1,1})$ varies between [0,1], representing the distribution of each player's production across all markets. Importantly, this function demonstrates concavity. The equilibrium is located at the maximum point, which occurs when $f(a_{1,1})=0.5$, as illustrated in Figure \ref{Utility_Graph}.
\color{black}

In the case where there are 2 players, the optimal strategy can be found analytically as shown in the next section, for different time horizons.

\subsection{Time Horizon Dependent Best Response}

Consider $t$ is the time step, $x_t$ is the strategy played by $x$ at $t$ and $a_t$ is the strategy played by the opponent at $t$. Equation \eqref{equ:utilitysmall} is drawn from $A_i(s_i,s_{-i})$ and the specific formulation given in Equation $\eqref{Equ:a}$ 
$$
u(x,t)=x_t(x_t+a_t)-x_t(-x_t-a_t)
$$
\begin{equation}
	u(x,t)=2x_t(x_t+a_t)
	\label{equ:utilitysmall}
\end{equation}

We shall now show that the best response is time horizon $\left(t_h\right)$ dependent and give proofs of it for time horizons $1,2$ and $\infty$.

The myopic best response is defined to be the strategy a player will choose if they consider only their utility for a time horizon $t_h=1$.

The strategy choices of the players are normalised, such that the equilibrium position is represented by 0 and all strategies are in $-1<s_i<+1$. A strategy $[s_1,s_2]=[-0.5,1]$ represents player 1 selling $\frac{3}{4}$ of their goods to market $1$ and $\frac{1}{4}$ to market 2 and player 2 selling all of their goods to market $2$.

\begin{thm}
	In the 2-player Multimarket Cournot Game with zero transportation costs and a utility defined in Equation \eqref{equ:utilitysmall}, the myopic best response to $a_0$ is $\frac{-a_0}{2}$, where $[x,a_0]$ is the set of strategies played by the players at $t=0$. 
	\label{Thm:time_horizon_1}
\end{thm}

The myopic best response considers only time horizon $t_h=1$. The utility equation for $i$ can then be calculated, and due to its concavity, the unique maximum will be the best response. The best response $\frac{-a_0}{2}$ is negative as after the normalisation, with the equilibrium strategy represented by 0, a best response represents recognising a player is selling too much to a particular market, and selling to the other instead. $[x,a_0]$ represents $(s_i,s_{-i})$, where $x$ is a strategy following $i's$.
\color{black}
\begin{pf}
	The set of strategies played by the players at $t=0$ (the only time step being evaluated) is $[x,a_0]$. Substituting into Equation \eqref{equ:utilitysmall} this is
	$$
	u(x,t)=-2x^2-2a_0x.
	$$
	This has a maximum where $\frac{\partial u(x,t)}{\partial x}=0$.
	$$
	\frac{\partial u(x,t)}{\partial x}=-4x-2a_0=0
	$$
	and so $x=\frac{-a_0}{2}$, confirming the result.
	\begin{flushright}
		$\square$
	\end{flushright}
\end{pf}


\begin{thm}
	It can be shown numerically that in the 2-player Multimarket Cournot Game with zero transportation costs and a utility defined in Equation \eqref{equ:utilitysmall}, the optimal response considering for a time horizon of 2 moves is $\frac{-a_0}{3}$, where $[x,a_0]$ is the set of strategies played by the players at $t=0$.
	\label{Thm:time_horizon_2}
\end{thm}
As in Theorem \ref{Thm:time_horizon_1}, the optimum can be found by examining the utility over time steps 1 and 2 due to the concavity of the utility function, the maximum will be the best response.
\begin{pf}
	Due to the symmetry of players strategies, if a player uses strategy $(x) (a_0)$, then their opponent will respond with $(x^2) (a_0)$. The utility over the two timesteps is therefore given by
	$$
	u(x,0)+u(x,1)=2a_0^2 x^2a_0^2 x + 2a_0^2  x^3 +2a_0^2 x^2
	$$
	$$
	u(x,0)+u(x,1)=2a_0^2 (x^3+2x^2+x)
	$$
	This has fixed points where $\frac{\partial (u(x,0)+u(x,1))}{\partial x}=0$.
	$$
	\frac{\partial (u(x,0)+u(x,1))}{\partial x}=2a_0^2  (3x^2+4x+1)=0
	$$
	has roots at $x=-1$ and $x=\frac{-1}{3}$. The root at $x=-1$ corresponds to both players repeatedly playing a market strategy equivalent to one another of the form $[-a_0,a_0]$. A strategy of $-a_0$ is a local minimum and players can improve their utility by moving away from it in either direction. For $x<-1$, the utility does increase. However this represents each player exponentially increasing the amount they sell to the market not dominated by their opponent, and recouping their losses when their opponent does the same. Due to the finite nature of markets, this is not a long term strategy and can be disregarded.
	
	The root at $x=\frac{-1}{3}$ is a local maximum and accordingly represents the best strategy for the time horizon of 2.
	
	\begin{flushright}
		$\square$
	\end{flushright}
\end{pf}

\begin{rem}
	In the 2-player Multimarket Cournot Game with zero transportation costs and a utility defined in Equation \eqref{equ:utilitysmall}, the optimal response considering for a time horizon of $1\leq t_h\leq 1000$ moves remains in the interval $\left[\frac{-a_0}{2},\frac{-a_0}{3}\right]$, where $[x,a_0]$ is the set of strategies played by the players at $t=0$.
	\label{Thm:time_horizon_3}
\end{rem}
As before, the optimum can be found by examining the utility over time steps $1-t_h$. Due to the concavity of the utility function, the maximum will be the best response. Figure~\ref{Fig:timehorizonmedium} shows this for $1<t_h<1000$. By referring to the basis in grain or oil transport, it can be observed that a reasonable duration for a time horizon is a month or longer. Where $t_h=1000$ is therefore more than $80$ years, longer than even the most long-sighted companies will optimise over. In fact, the following theorem confirms this numerically for the limit case $t_h\rightarrow \infty$.

\begin{thm}
	In the 2-player Multimarket Cournot Game with zero transportation costs and a utility defined in Equation \eqref{equ:utilitysmall}, the optimal response considering for an infinite time horizon is $\frac{-a_0}{1+\sqrt(2)}$, where $[x,a_0]$ is the set of strategies played by the players at $t=0$.
	\label{Thm:time_horizon_4}
\end{thm}
As in Theorems \ref{Thm:time_horizon_1} and \ref{Thm:time_horizon_2}, the optimum can be found by examining the utility over all time steps due to the concavity of the utility function, the maximum will be the best response. 
\begin{pf}
	For large $t_h$, the structure of the utility function remains the same and is given by $u$. The fixed point therefore occurs at $\frac{\partial u}{\partial x}=0$.
	$$
	\frac{\partial u}{\partial x}=\frac{\partial (u(x,0)+u(x,1)+u(x,2)+\dots+u(x,t_h))}{\partial x}=0
	$$
	\begin{equation}
		\frac{\partial u}{\partial x}=(t_h+1)*x^{t_h}+\left(\sum_{i=1}^{t_h-1} 2(i+1) x^{i}\right) +1 =0.
		\label{Equ:TimeHorizonInfty}
	\end{equation}
	This is the sum of successive terms of \ref{equ:utilitysmall} for different time steps, taking into consideration that when a player updates their strategy, all future time steps are affected by that update.
	We can use the identity 
	\begin{equation}
		\frac{1}{(1-x)^2}=\sum_{n=1}^{\infty} \left(n x^{(n-1)}\right)
		\label{Equ:InfiniteSum}
	\end{equation}
	By substituting Equation \eqref{Equ:InfiniteSum} into Equation \eqref{Equ:TimeHorizonInfty}, we get
	$$
	\frac{\partial u}{\partial x}=2\sum_{n=1}^{\infty} \left(n x^{(n-1)}\right) -2+1=0
	$$
	$$
	\frac{\partial u}{\partial x}=\frac{2}{(1-x)^2}-2+1=0.
	$$
	A fixed point where $\frac{\partial u}{\partial x}=0$ occurs at $x=\frac{-1}{1\pm\sqrt(2)}$. Of these, we need only consider the negative case, $x=\frac{-1}{1+\sqrt(2)}$ as the alternate solution results in players receiving negative utility relative to the fixed case at all time steps. Accordingly, a response $x a_0$ where $x=\frac{-1}{1+\sqrt(2)}$ is optimal for an infinite time horizon.
	\begin{flushright}
		$\square$
	\end{flushright}
\end{pf}


We now introduce the equilibrium mimicking strategy and define it as
\begin{definition}
	An equilibrium-mimicking strategy profile is a set of strategies played by all players such that all markets receive the same supply as they would if the game was in equilibrium. 
\end{definition}

\begin{thm}
	In the two-player two-market game, let $A_i(\cdot,\cdot)$ be the utility received by the first player in the upper layer as defined in Equation \eqref{Equ:A2} and let $g_i\geq 0$ be the difference in utility gained after the first move as a myopic best response, compared to the utility the first player would receive if they played an equilibrium mimicking strategy. Then the utility of the first player through successive myopic best responses, compared to the final unique equilibrium strategy is 
	$
	\frac{8g_i}{15}>0.
	$
\end{thm}
This implies that rational players will have no incentive to play an equilibrium strategy if all the other players have not already done so.
\begin{pf}
	Normalising with respect to an equilibrium mimicking strategy wherein players consistently play $[0,0]$, with $g_i$ the additional utility gained by the myopic responder relative to the $[0,0]$, the utility gained by the player at each subsequent change in strategy by either player is given by the sequence
	$
	g_i,-\frac{g_i}{2},\frac{g_i}{16},-\frac{g_i}{32},\dots
	$
	
	This is an infinite sequence expressed as
	$$
	\sum_{n=1}^\infty s_n=\frac{s_1}{1-(\frac{s_2}{s_1})},
	$$
	sums to
	$$
	\frac{8g_i}{15}>0.
	$$
	\begin{flushright}
		\qed
	\end{flushright}
\end{pf}

\begin{figure}[h]
	\centering
	\begin{minipage}{0.6\textwidth}
		\centerline{\includegraphics[width=\columnwidth]{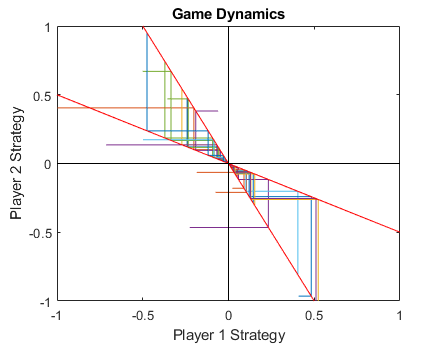}}
		\caption{The Location of the Strategy for discrete time steps from a variety of starting points.}
		\label{Game_Dynamics}
	\end{minipage}\hfill
\end{figure}

Consequently, each player adopts a myopic best-response strategy based on their opponent's current strategy, instead of immediately adhering to the equilibrium strategy. The myopic best response proves to be a superior strategy across all time frames. As a result, the dynamics of the game gradually approach equilibrium, as illustrated in Figure \ref{Game_Dynamics}. The bounds on convergence are highlighted in red, and they become applicable after just two responses.

\begin{thm}
	In the two-player game, let $A_i(\cdot,\cdot)$ be the utility received by the first player in the upper layer be as defined in Equation \eqref{Equ:A2} and let $g_i\geq 0$ be the difference in utility gained after the first move for an infinite time horizon optimal strategy (as defined in Theorem \ref{Thm:time_horizon_4}), compared to the utility the first player would receive at equilibrium. Then the utility of the first player through successive infinite time horizon optimal moves, compared to the final unique equilibrium strategy is 
	$$
	\frac{g_i(1+\sqrt(2))}{4}>0.
	$$
\end{thm}

\begin{pf}
	The sequence $s_i$ of utilities received by the first player (normalised with respect to the equilibrium utility) is $g_i,(1-\sqrt(2)) g_i,(1-\sqrt(2))^4 g_i,(1-\sqrt(2))^5 g_i,\dots$.
	
	\noindent Using the formula
	$$
	\sum_{n=1}^\infty s_n=\frac{s_1}{1-(\frac{s_2}{s_1})},
	$$
	it is found that the sequence sums to
	$\frac{g_i (1+\sqrt(2))}{4}$.
	
	\begin{flushright}
		\qed
	\end{flushright}
\end{pf}

It can be seen that, as expected, $\frac{1+\sqrt{2}}{4}>\frac{8}{15}$, demonstrating the infinite time horizon pattern results in a higher utility for the player than the myopic strategy.

\begin{thm}
	In the 2-player Multimarket Cournot Game with zero transportation costs and a utility defined in Equation \eqref{equ:utilitysmall}, the optimal response where each timestep is given weight proportional to $\beta^t$ for $\beta\in [0,1]$ is given by 
	$$
	\omega=\frac{\beta+1-\sqrt{\beta^2-\beta+1}}{3\beta}.
	$$
	where $\omega$ is the constant representing the magnitude of the response, such that if $S$ is given by $[-,a_0]$, the response is given by $[-\omega a_0,a_0]$.
\end{thm}
$$
V(x_0)=\max_{\{S_i^t\}_{t=0}^\infty}  \sum_{t=0}^\infty \beta_t F(S^t,S_i^t)
$$
is the Bellman equation \cite{Bellman} with $F(S^t,S_i^{t})$ the utility given by the players' strategy update rule and $S^t$ is the strategy profile at time t. The strategy update rule assumes that at each time $t$, the active player chooses the best response over all times, $s_i^{*,t}$ from the set $S_i^{t}$.
\begin{pf}
	Utility is given by the current player's response $\omega$ to the current state of the system. The utility given by the response from the next two time steps is given by
	$$
	u(x,1)=(c)(1-c)(a)(c-c^2)(a^2)
	$$
	$$
	u(x,2)=\beta(c)(c^2-c)(a^2)
	$$
	The best response is given at the maximum of $G=u(x,1)+u(x,2)$. This is found at the maximum of G, which occurs where $\frac{dG}{dc}=3\beta c^2 -2(\beta+1)c+1=0$. This has two maxima. The positive maxima represents a solution which requires progressively more extreme responses to their market, which given the finite nature of trade goods is , and accordingly, the maximum is given at $\frac{\beta+1-\sqrt{\beta^2-\beta+1}}{3\beta}.$
\end{pf}

\subsection{Characteristics under Player Collusion}

An assumption made in the classical Cournot problem is the absence of collusion between players, and each chooses their strategy independently. A group of players engaging in collusion is described as a `coalition'. Members of a coalition improve their utility by rejecting strategies that are not pareto-optimal for the subset of player in the coalition, in favour of acting to mutually increase their utilities. 

We shall distinguish between `equilibrium play' and  `advantage play'.  `Equilibrium play' is the specific strategy which players immediately move to the strategy they would play in the final equilibrium. This fixes their utility preventing them from making both losses and profits relative to this fixed amount. In comparison, `advantage play' is any other strategy in which players attempt to profit from the current out of equilibrium state. 

Where players work together, equilibrium play is never optimal.
To prove that, we first define a Pareto-optimal strategy as follows
\begin{definition}
	A strategy profile $s^P$ is a pareto optimum if for all $s\neq s^p$ in $S$, there exists $i$ such that 
	$$
	u_i(s)< u_i(s^p).
	$$
\end{definition}
\begin{thm}
	At any player count $N$ where the starting conditions are not at equilibrium, there exists a pareto optimum strategy where equilibrium play is dominated by advantage play.
\end{thm}
\begin{pf}
	Assume towards a contradiction that there is no pareto-optimal strategy wherein all players benefit from engaging in advantage play. Accordingly, all players will play an equilibrium strategy. 
	Consider player $i$. The system is not currently at equilibrium and accordingly $\frac{\partial c_i}{\partial x_i}\neq 0$. As such, either $x_i=+\epsilon$ or $x_i=-\epsilon$ results in a cost $c_i < c$. If each player follows this strategy all players would have costs less than $c$. This would mean that every player received positive utility relative to the equilibrium case and accordingly players are incentivised to cooperate to bring this about. As such, the equilibrium strategy can be rejected as it is dominated.
\end{pf}

\subsection{Lower Layer Dynamics - Approach to Transport Equilibrium}

Concentrating on the lower, consider the upper layer transport dynamics to be fixed such that there is a fixed set of flow to be allocated between a series of origin-destination pairs through the lower layer.


Total Transport is defined to be the amount of goods a player has available to allocate between all markets they have access to.

Consider a transport problem as a congestion game such that, in its simplest form, the lower layer has a two path structure, where there are two symmetric paths, $e_1$ and $e_2$ from $v_1$ to $v_2$. There are two routes from the origin to the destination, and $N$ players attempting to minimise their transport costs with fixed total transport quantity $T_i$ associated with each player, where $T_i$ is a fixed constant such that
$$
\sum_{j=1}^{2} f(x_i,e_j)=T_i
$$ 
for all $i$ in ${1,\dots,N}$. As such, the seller's output is bounded but the market inputs are unbounded. The strategy of players in the upper layer is fixed and so there is no feedback loop from the outcome of the transport problem to the upper layer. First consider the case in which the cost to player $x_i$ for using edge $e_j$ is given by 
\begin{equation}
	f(x_i,e_j) \sum_{k=1}^N f(x_k,e_j)
\end{equation}
where $f(x_i,e_j)$ is the flow of player $x_i$ on edge $e_j$. This is the same structure (without costs) as is introduced in section 2 of \cite{Bimpikis}. Quadratic costs are a common assumption in the literature \cite{Vives}, \cite{Singh}. Where all other players strategies are fixed, the cost to player $i$ for using edge $e_j$ is given by 
$$
f(x_i,e_j)(c+f(x_i,e_j))
$$
where $c$ is a constant representing the choices of all other players.

The assumption is that each player has a fixed total transport $T_i$. Accordingly,
$$
\sum_{j=1}^N f(x_i,e_j)=T_i \mbox{ for all } i \in \{1,\dots,N_x\}.
$$

Therefore, the set of strategies in $s_{i}$ has one degree of freedom as
$$f(x_i,e_2)=T_i-f(x_i,e_1).$$
As $T_i$ is fixed, $f(x_i,e_2)=T_i-f(x_i,e_1)$ and the strategy space $s_i$ available to player $i$ has only one degree of freedom. As such it can be represented by a single value. We set this value to be the amount of transport sent along route $e_1$. For generality, $s_{i} \in [-\frac{\tau}{2},\frac{\tau}{2}]$. This is normalised such that $s_{i}=-\frac{\tau}{2}$ representing a pure strategy in which $f_{\cdot,\cdot}=0$, $s_{i}=\frac{\tau}{2}$ corresponding to the pure strategy in which $f(x_1,e_1)=T$ with all mixed strategies corresponding to $s_i\in(0,1)$


Considering a congestion function $B(f(\cdot))$ which increases monotonically with flow $f(\cdot)$, we can now state the following Theorem
\begin{thm}
	\label{thm4}
	Consider two values of the flows  $f(e_1^\alpha)$ and $f(e_2^\alpha)$ between any two nodes in the upper layer, with  $f(e_1^\alpha)>f(e_2^\alpha)$.  When $B(f(e_1^\alpha))\geq B(f(e_2^\alpha))$ \textit{ and } $w(a_{i,j})\in \mathbb{R_+}$, then \textit{ Wardrop Equilibria exist in the lower layer}.
\end{thm}
In fact, where the transportation costs are monotonically increasing as flow increases and assuming fixed edge weights in the upper layer, Wardrop equilibria can be found in \cite{Wardrop}.

\begin{pf}
	Consider the rate of change of utility with respect to changes made in a player's transport strategy, given by
	\begin{equation}
		\frac{\partial u(s_i,s_{-i})}{\partial p_{i,j}^k}
	\end{equation}
	When considering a change in the paths selected a player can examine the marginal change in congestion cost from doing so. Increasing the amount the player uses a path will always result in an increase in costs. Due to the fixed upper layer however, $w(a_{i,j})$ is constant and
	
	\begin{equation}
		\sum_{k=1}^{N_{P_{i,j}}} p_{i,j}^k = w(a_{i,j}) 
	\end{equation}
	is fixed. As such any additional traffic along one path will result in a proportional reduction in traffic along a parallel path or paths. Parallel paths can share edges, but no path can use all the edges used by another path with the same origin and destination. The path allocation is therefore in equilibrium when 
	\begin{equation}
		\frac{\partial u(s_i,s_{-i})}{\partial p_{ij}^k}=q_{ij}, q_{ij}\in \mathbb{R}_+, p_{ij}^k\neq 0 \mbox{ for all } i,j,k.
	\end{equation}
	
	This results in all paths available to a player having the same cost, fulfilling the conditions of a Wardrop equilibrium.
\end{pf}
\begin{flushright}
	\qed
\end{flushright}

The inter-layer connections provide a feedback loop that so far has not been considered. It is important to note that for Theorem \ref{thm4}, paths must exist in $G_\beta$ between $L(x_i^\alpha)$ and $L(y_j^\alpha)$ for which  $w(a_{i,j})>0$. In the complete model however, when there does not exist a path between $L(x_i^\alpha)$ and $L(y_j^\alpha)$, the transport layer would give infinite congestion cost for trade between $x_i$ and $y_j$. As such they would trade between this player and market in the upper layer would stop. The players would then always have an incentive to change to another strategy in the upper layer creating a beneficial feedback loop wherein players only sell to markets they can reach.

\subsection{Transport costs over multiple paths}

Transport costs relate to edges and for each player, the cost relates to the share of transport over a number of paths. As the players allocate transport to different paths, the combined transport cost loses differentiability. This allows us to state the following theorems. In the case of Theorem \ref{thmdifferentiability} each player wants to use different paths for which the individual transport cost is continuous and twice differentiable, the result over all paths is continuous and differentiable, but not continuous and twice differentiable.

\begin{thm}
	\label{thmdifferentiability}
	Transportation costs are continuous and differentiable, but not twice differentiable.
\end{thm}

\begin{pf}
	The non-differentiability follows from the switching from one path to the other as congestion costs interfere on the former. A full proof can be found in the appendix.
\end{pf}

\color{blue}
\color{black}

\section{General Multilayer Solutions}

When examining the interconnected layers, it's important to note that the proof of the existence of unique equilibria, as presented in \cite{Bimpikis} and utilized in section \ref{Upperlayeronly}, is not applicable. This is due to the fact that, while the profit functions maintain their properties of being twice differentiable, convex, and strictly decreasing, the cost functions do not exhibit these characteristics; instead, they lack second differentiability, are concave, and exhibit an increasing trend. However, we can establish the necessary and sufficient condition for equilibrium in the multilayer game through the following theorem.

\begin{thm}
	The Cournot-congestion game achieves equilibrium if and only if, for each player,
	\begin{equation}
		\frac{\partial u(s_i,s_{-i})}{\partial a_{ij}}-\frac{\partial u(s_i,s_{-i})}{\partial p_{ij}^{k}}=r(i) \mbox{  } \forall p_{ij}^k\neq 0
	\end{equation}
	for $i \in {1,2,\dots,N}$, $j \in {1,2,\dots,M}$, $k\in {1,2,\dots,N_{P_{ij}}}$, and $r(i)\in\mathbb{R}_+$, with $r(i)$ representing a fixed constant for all $j$ and $k$.
\end{thm}

This implies that the system is in equilibrium if and only if each player possesses the same marginal utility for all (non-zero) strategies available to them. This is a unique and stable equilibrium.

\begin{pf}
	The change in utility for a market and its associated route is given by
	\begin{equation}
		\frac{\partial u(s_i,s_{-i})}{\partial a_{ij}}-\frac{\partial u(s_i,s_{-i})}{\partial p_{ij}^{k}}=r(i) \mbox{  } \forall p_{ij}^k\neq 0
		\label{Equ:GeneralEquilibrium2}
	\end{equation}
	If there are no alternative markets or paths that offer a utility increase greater in magnitude than the decrease in utility from not using the current strategy, a player will consider changing strategies. This implies that for any $s \in S$ for which Equation \eqref{Equ:GeneralEquilibrium2} is not valid, the system is not in equilibrium.
	
	Conversely, if the system were in equilibrium but a player had varying marginal utility for different strategies, that player would have an incentive to adjust their mixed strategy profile to favour strategies with higher marginal utility. This incentive for strategy change would disrupt the equilibrium. Thus, if all players have the same marginal utility for each strategy, no player has an advantage in making marginal changes to their strategy. Consequently, the system must be in equilibrium.
\end{pf}
\begin{flushright}
	\qed
\end{flushright}

\begin{corollary}
	The strategy profile $S^{E}$ which meets the conditions outlined in equation~\ref{Equ:GeneralEquilibrium2} is a unique and stable equilibrium.
\end{corollary}

\begin{pf}
	The stability property is a result of the utility functions for both the upper and lower layer having lyapunov properties. This results in a dampening effect on deviation from the equilibrium.
\end{pf}
\begin{flushright}
	\qed
\end{flushright}


\section{Case Study}

\subsection{3 Player 3 Market Case}

\begin{figure}[h]
	\begin{minipage}{0.45\textwidth}
		\centerline{\includegraphics[width=\columnwidth]{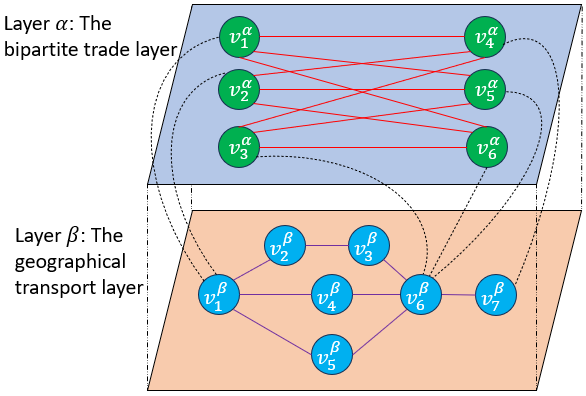}}
		\caption{The Multilayer network for the case study.}
		\label{CSGRAPH}
	\end{minipage}
	\centering
	\begin{minipage}{0.45\textwidth}
		\centerline{\includegraphics[width=\columnwidth]{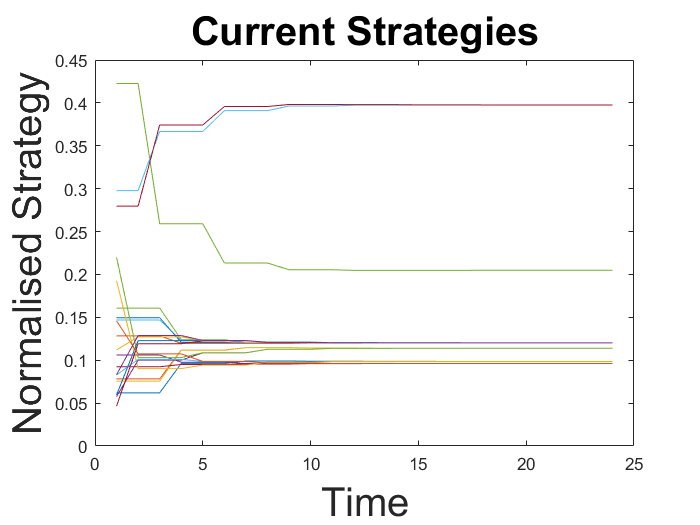}}
		\caption{Time against Strategy for each of the 21 paths}
		\label{CSSTRATEGIES}
	\end{minipage}
	\begin{minipage}{0.45\textwidth}
		\centering
		\centerline{\includegraphics[width=\columnwidth]{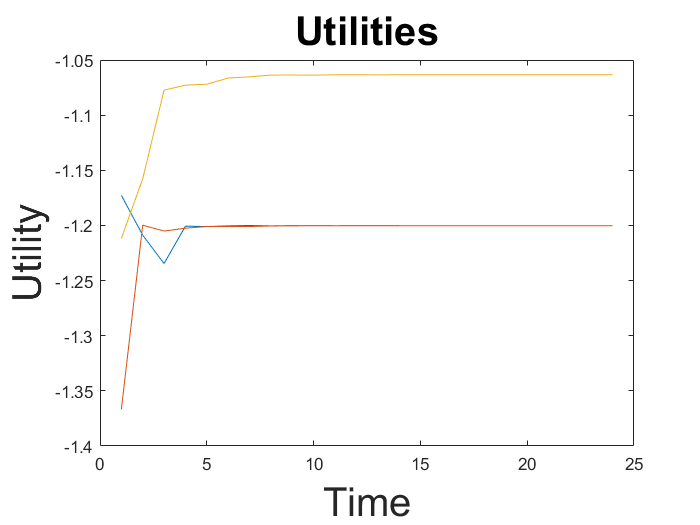}}
		\caption{Time Against Utility for each of the three players}
		\label{CSUTILITIES}
	\end{minipage}\hfill
	\begin{minipage}{0.45\textwidth}
		\centerline{\includegraphics[width=\columnwidth]{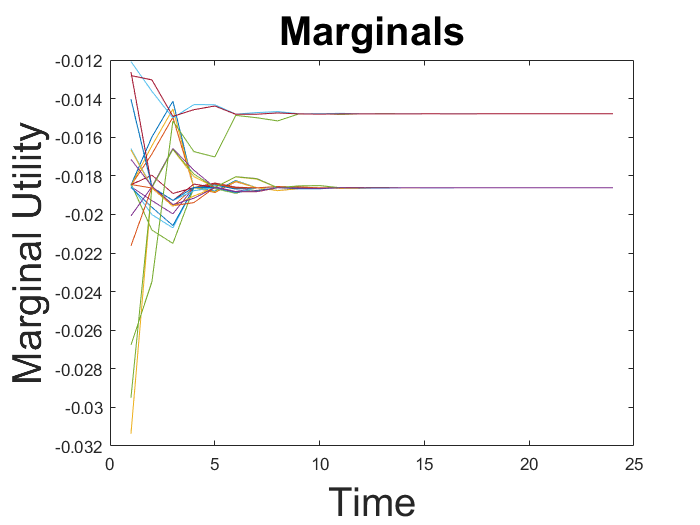}}
		\caption{Time against Marginal Utility of each of the 21 paths}
		\label{CSMARGINALS}
	\end{minipage}
\end{figure}

Consider the multilayer network given in figure \ref{CSGRAPH}. There are 3 players and 3 markets. Players 1 and 2 are located in node 1 of the lower layer, and player 3 is located in node 6 of the lower layer. Market 1 is located in node 7 of the lower layer, and markets 2 and 3 are located in node 6.

The equilibrium of this system is shown in figure~\ref{CSSTRATEGIES}. It can be seen in figure~\ref{CSUTILITIES} that the marginal utilities of each player converge to a stable position. The utilities of players 1 and 2 have the same equilibrium position due to their symmetry. Finally it can be seen that the two symmetric players receive the same normalised utility, while player 3, who is located closer to the markets, receives a higher utility, as shown in figure~\ref{CSMARGINALS}.


\color{black}

\section{Discussion}

This paper examines the dynamics of two interconnected games that collectively capture behaviours observed in global trade. The strategic interactions of players in oligopolistic settings are effectively encapsulated by the Cournot competition, a concept that has been studied since the early $20^{th}$ century. Similarly, congestion games, which pertain to strategic traffic routing, represent a classical approach within operations research for addressing transportation problems, with well-established results concerning the Wardrop equilibrium.

However, when these two games are situated on graphs and intertwined within a multilayer network, the dynamics become notably more intricate. In this context, equilibria within each layer are influenced by the structure and dynamics of the other, yielding novel insights. 

The choice to represent these intertwined systems through a multilayer model requires justification. Where there exist congestion functions on both layers, a system of multilayer networks can be reduced to a single layer, where it becomes an origin-destination network problem. Where the bipartite network of markets in layer $\alpha$ is complete (such that every player has access to every market), this can be done without constraints, and the problem is fully reducible. Where the bipartite network of markets is not fully connected, it requires path constraints. The general reducibility of multilayer networks is supported by \cite{DeDomenico2015}. However, an example appears in Appendix~\ref{Section:GraphReduction} of two different multilayer networks of combined trade and transport which would be reduced to the same single layer network. This represents a case where reduction to a single layer results in a loss of information, and accordingly the multilayer structure is appropriate.

A previous attempt to introduce transportation dynamics as a cost within the Cournot competition was made by \cite{Bimpikis}, where each player-market interaction was associated with a congestion cost contingent upon the flow of goods along the corresponding edge. By segregating the transport layer from the Cournot competition, our work introduces the possibility of multiple routing paths. Consequently, a player's strategy encompasses not only the quantity of goods but also the choice of routing to the market.

While travel costs were previously incorporated into the Cournot competition by \cite{Alsabah2021}, their approach focused solely on functional transport costs. The multilayer, path-based transport routing introduced in this paper represents a unique contribution to the existing literature.

By analysing the two layers separately, our findings align with prior research on single-layer network oligopolies and selfish routing problems, which we leverage in our analysis. However, a key distinction arises between the fixed travel dynamics discussed in Section \ref{Upperlayeronly} and the classical Cournot competition. In our single-layer Cournot game analysis, transportation costs still exist, but players are unable to influence them through the choice of alternative routing. Our results shed light on the nature and rate of convergence toward equilibrium. It is found that while the multilayer network is in equilibrium, the entangled nature of the two layers can result in neither of them being in equilibrium when considered in isolation.

Under mild assumptions regarding the convexity and concavity of costs and profits, our model offers valuable insights into the benefits of adopting a selfish strategy as equilibrium is approached. While this has been proven for the case of two players and any number of markets, an extension to multiple players is currently under investigation.


\section{Conclusions}

In this study, a multilayer network framework was employed to establish a connection between the network Cournot competition (upper layer) and a selfish routing problem (lower layer). When examining the behaviour of players in the upper layer in isolation from the lower layer, it was observed that a unique stable equilibrium emerged. Similarly, an investigation of the lower layer revealed the presence of a unique stable equilibrium.

However, when these layers were coupled together, the existence of unique stable equilibria no longer held. This was due to the concave behaviour exhibited in the upper layer and the convex behaviour seen in the lower layer, which, when combined, resulted in a non-concave utility function. While the existence of equilibria was proven to exist within this coupled framework, the dynamics of these equilibria were explored using a link-route representation of the lower layer.
\color{black}

	In this work, a multilayer network was used to couple the network Cournot competition (on the upper layer) to a selfish routing problem (on the lower layer). The nature of the behaviour of players on the upper layer when uncoupled from the lower layer has been found to manifest as a unique stable equilibrium. Similarly it was found that there exists a unique stable equilibrium on the lower layer. When the layers are coupled there no longer exists unique stable equilibria as the concave behaviour in the upper layer and the convex behaviour in the lower layer when added give a  non-concave utility function. 
\color{black}

\section*{Declarations}

\subsection*{Ethical Approval}

Not applicable.

\subsection*{Competing Interests}

Toby Willis declares that he has no conflict of interest. Giuliano Punzo declares that he has no conflict of interest.

\subsection*{Authors' Contributions}

The authors equally contributed to this work.

\subsection*{Funding}

This research did not receive funding.

\subsection*{Availability of Data and Materials}

Not applicable.

\bibliography{sn-bibliography}

\appendix
\section{Proofs}    
\begin{nthm}
	Transportation costs are continuous and differentiable, but not twice differentiable.
\end{nthm}
\begin{pf}
	Consider a player attempting to minimise transportation costs from $L(x_i)$ to $L(y_j)$ across 3 three routes. The flows on the routes are given by $x,y$ and $z$ with the cost on each route given by $(x)(x+a)$, $(y)(y+b)$ and $(z)(z+c)$ and without loss of generality $a\leq b \leq c$ . The total transportation cost is then given by
	$$
	t:\mathbb{R}^3\rightarrow\mathbb{R}.
	$$
	The total flow which $x_i$ routes between $L(x_i)$ and $L(y_j)$ is $f$ and as such $x+y+z=f$. The minimum of $t(x,y,z)$ is given by $T(f)$.
	$$
	T:\mathbb{R}\rightarrow\mathbb{R}.
	$$
	The marginal rate of change of $t$ with respect to each edge are given by:
	$$
	\frac{\partial t}{\partial x}=2x+a, \frac{\partial t}{\partial y}=2y+b, \frac{\partial t}{\partial z}=2z+c.
	$$
	As such, there are changes of behaviour of $T(f)$ at $f=\frac{b-a}{2}$ and at $f=\frac{2c-a-b}{2}$.
	\begin{equation}
		T(f)=\begin{cases}
			t(f,0,0) &\mbox{for} f < \frac{b-a}{2}\\
			t\left(\frac{f}{2}+\frac{b-a}{4},\frac{f}{2}-\frac{b-a}{4},0\right) & \mbox{for} \frac{b-a}{2} < f < \frac{2c-a-b}{2}\\
			t\left(\frac{f}{3}+\frac{-2a+b+c}{6},\frac{f}{3}+\frac{a-2b+c}{6},\frac{f}{3}-\frac{2c-a-b}{6}\right) & \mbox{for} f > \frac{2c-a-b}{2}.
		\end{cases}
	\end{equation}
	This is continuous as
	$$
	\mbox{lim}^-_{\frac{b-a}{2}} T(f) = \frac{b^2-a^2}{4} = \mbox{lim}^+_{\frac{b-a}{2}} T(f)
	$$
	and
	$$
	\mbox{lim}^-_{\frac{2c-a-b}{2}} T(f) = \frac{-a^2-b^2+2c^2}{4} = \mbox{lim}^+_{\frac{2c-a-b}{2}} T(f).
	$$
	The first derivative is given by
	\begin{equation}
		\frac{d T}{df}=\begin{cases}
			2f+a &\mbox{for} f < \frac{b-a}{2}\\
			f+\frac{a+b}{2} & \mbox{for} \frac{b-a}{2} < f < \frac{2c-a-b}{2}\\
			\frac{2f}{3}+\frac{a+b+c}{3} & \mbox{for} f > \frac{2c-a-b}{2}.
		\end{cases}
	\end{equation}
	$T$ is differentiable as
	$$
	\mbox{lim}^-_{\frac{b-a}{2}} \frac{dT}{df} = b = \mbox{lim}^+_{\frac{b-a}{2}} \frac{dT}{df}
	$$
	and
	$$
	\mbox{lim}^-_{\frac{2c-a-b}{2}} \frac{dT}{df} = c = \mbox{lim}^+_{\frac{2c-a-b}{2}} \frac{dT}{df}.
	$$
	The second derivative is given by
	\begin{equation}
		\frac{d^2 T}{df^2}=\begin{cases}
			2 &\mbox{for} f < \frac{b-a}{2}\\
			1 & \mbox{for} \frac{b-a}{2} < f < \frac{2c-a-b}{2}\\
			\frac{2}{3} & \mbox{for} f > \frac{2c-a-b}{2}.
		\end{cases}	
	\end{equation}
	$\frac{d^2 T}{d f^2}$ is not continuous at $\frac{b-a}{2}$ and $\frac{2c-a-b}{2}$ and accordingly $T$ is not twice differentiable.
	\begin{flushright}
		$\square$
	\end{flushright}
	It has been shown that $T$ is not twice differentiable at the transition between assigning additional flow to 1 path and assigning it to 2 paths and also that $T$ is not twice differentiable at the transition between assigning additional flow to 2 path and assigning it to 3 paths. By the same calculations, it can be seen that $T$ is not twice-differentiable at the change in behaviour between $k$ and $k+1$ paths being used. Between $k$ paths and $k+1$ paths, $T$ will hold the following behaviour
	\begin{equation}
		T(f)=\begin{cases}
			t\left(\frac{f}{k}+\frac{-(k-1)a+b+\dots+l_k}{2k},\frac{f}{k}+\frac{a-(k-1)b+\dots+l_k}{2k}\dots,0\right) & \mbox{for} \frac{2l_k-a-b-\dots l_{k-1}}{2} < f < \frac{2l_{k+1}-a-b-\dots-l_k}{2}\\
			t\left(\frac{f}{k+1}+\frac{-(k)a+b+\dots+l_{k+1}}{2k+2},\frac{f}{k+1}+\frac{a-(k)b+\dots+l_{k+1}}{2k+2}\dots\right) & \mbox{for} f > \frac{2l_{k+1}-a-b-\dots-l_k}{2}.
		\end{cases}
	\end{equation}
	where $l_k$ is the $k^{th}$ letter of the alphabet (such that if $k=3$, $l_k=c$ and $l_{k+1}=d$).
\end{pf}

\section{Graph Reduction Example}
\label{Section:GraphReduction}

\begin{figure}[h]
	\centering
	\begin{minipage}{0.45\textwidth}
		\centering
		\centerline{\includegraphics[width=\columnwidth]{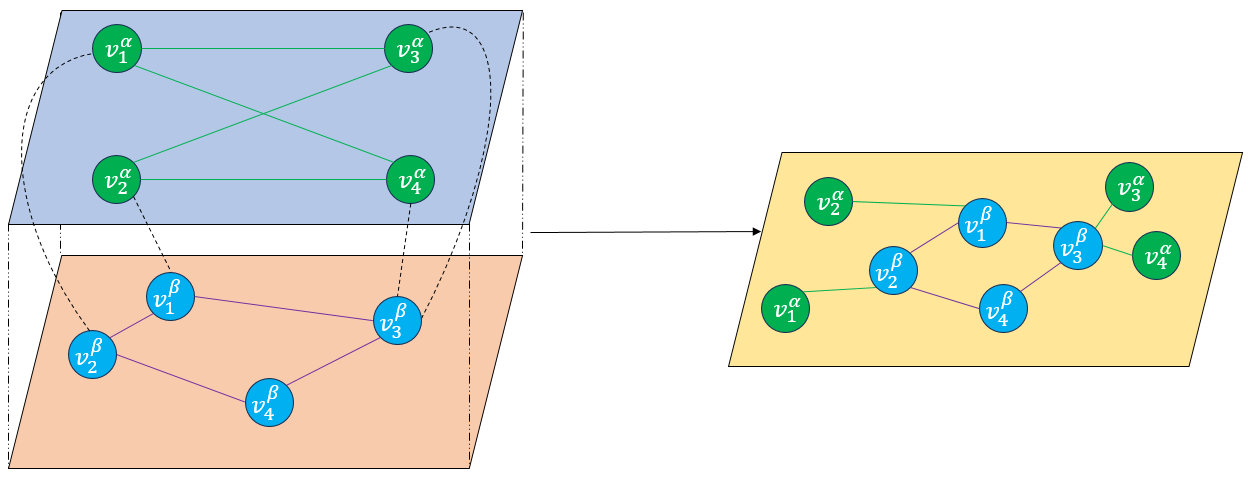}}
		\caption{The reduction to a single layer of a graph with a complete bipartite upper layer}
		\label{Figure:GraphReduction1}
	\end{minipage}\hfill
	\begin{minipage}{0.45\textwidth}
		\centerline{\includegraphics[width=\columnwidth]{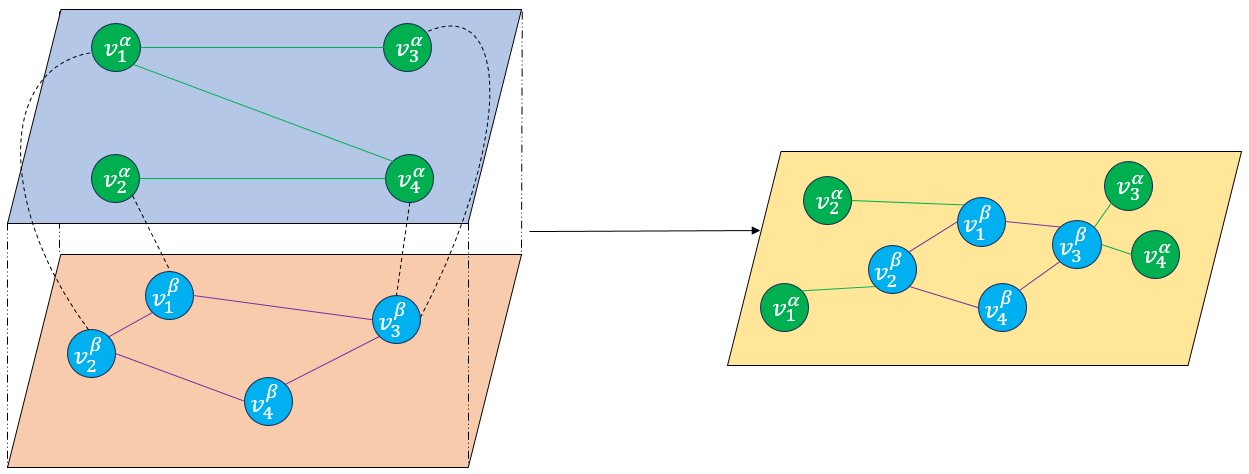}}
		\caption{The reduction to a single layer of a graph with an incomplete bipartite upper layer}
		\label{Figure:GraphReduction2}
	\end{minipage}
\end{figure}

Figures~\ref{Figure:GraphReduction1}~and~\ref{Figure:GraphReduction2} shows two separate multilayer networks, and the single layer graph that they would be reduced to by a graph reduction process.

This retains the edges connecting the players to the transport network, the edges within the transport network and the edges between the transport network and the markets. These edges are given congestion functions associated with their characteristics. The edges between the players and the transport networks are given 0 cost, and the edges between the transport networks and the markets are given congestion functions representing the market saturation. However, it can be seen that both Figure~\ref{Figure:GraphReduction1} and Figure~\ref{Figure:GraphReduction2} reduce to the same single layer graph. The information about which markets each player has access to is lost, and would require additional information to be given. Accordingly, reduction to a single layer graph is not possible without the loss of information, and a multilayer format is an appropriate choice.

\end{document}